\documentclass[english,aps,pra,superscriptaddress,nofootinbib]{revtex4}
\usepackage[T1]{fontenc}
\usepackage[latin9]{inputenc}
\usepackage{amsmath}

\makeatletter
\@ifundefined{textcolor}{}
{%
 \definecolor{BLACK}{gray}{0}
 \definecolor{WHITE}{gray}{1}
 \definecolor{RED}{rgb}{1,0,0}
 \definecolor{GREEN}{rgb}{0,1,0}
 \definecolor{BLUE}{rgb}{0,0,1}
 \definecolor{CYAN}{cmyk}{1,0,0,0}
 \definecolor{MAGENTA}{cmyk}{0,1,0,0}
 \definecolor{YELLOW}{cmyk}{0,0,1,0}
}


\newcommand{\1}{{\rm 1\hspace{-0.9mm}l}}

\makeatother

\usepackage{babel}
\begin{document}

\title{Uncertainty\textendash Reality Complementarity and Entropic Uncertainty
Relations}

\author{\L ukasz Rudnicki}
\email{rudnicki@cft.edu.pl}

\affiliation{Max-Planck-Institut f{\"u}r die Physik des Lichts, Staudtstra{\ss}e 2, 91058
Erlangen, Germany}

\affiliation{Center for Theoretical Physics, Polish Academy of Sciences, Al.
Lotnik{\'o}w 32/46, 02-668 Warsaw, Poland}
\begin{abstract}
Reality of quantum observables, a feature of long-standing interest
within foundations of quantum mechanics, has recently been quantified
and deeply studied by means of entropic measures {[}Phys. Rev. A \textbf{97},
022107 (2018){]}. However, there is no state-independent ''reality
trade-off'' between non-commuting observables, as in certain systems
all observables are real {[}Europhys. Lett. \textbf{112}, 40005 (2015){]}.
We show that the entropic uncertainty relation in the presence of
quantum memory {[}Nature Phys. \textbf{6}, 659 (2010){]} perfectly
supplements the discussed notion of reality, rendering trade-offs between reality and quantum uncertainty. State-independent complementarity
inequalities involving entropic measures of both, uncertainty and
reality, for two observables are presented.
\end{abstract}
\maketitle
\section{Introduction}

Shannon's information theory \cite{Shannon}, established 70 years
ago and celebrated in the current special volume, has influenced physics
across various domains. The range of applications for concepts and
tools developed by Shannon is extraordinarily broad, likely because
``\emph{the information described by Shannon's theory and measured
by the Shannon entropy is not classical, but is neutral with respect
to the physical theory that describes the systems used for its implementation}''
\cite{Filozofia}. As a consequence, several foundational and practical
aspects of quantum mechanics, subsumed under the popular name of quantum
information  have been studied with the help of Shannon entropy and its relatives.

My favorite aspect of quantum mechanics, in which the Shannon entropy has 
brought a new quality, is the field of uncertainty relations (URs).
Entropic uncertainty relations, pioneered in 1975 \cite{BBM} for
canonically conjugate pair of position and momentum, are a paradigm
shift in the theory of URs. It is because values assumed by observables
under discussion no longer play a role, instead, everything can be
efficiently characterized by means of sole probability distributions.
This conceptual novelty was soon adapted to the finite-dimensional
(discrete) scenario \cite{Deutsch} and substantially upgraded by
the celebrated Maassen-Uffink (MU) bound \cite{MU}. Recently, several
improvements (some on the conceptual side \cite{Ozawa}) of the MU
bound have been developed \cite{SRuiz,my1,Oni,CP,my2,KamilK,Zozor,Kaniewski,my3}
\textemdash{} see also \cite{Wehner2} and previous reviews on the
topic \cite{Wehner,Sen}. Let me mention in passing that ``in between''
both cases (continuous and finite-dimensional) there are, so-called,
coarse-grained observables for which the notion of uncertainty has
also been well-captured with the Shannon \cite{Partovi,IBB1984} and
R{\'e}nyi\footnote{The R{\'e}nyi entropy is the most important generalization of the Shannon
entropy, which appears after relaxation of one from the axioms originally
assumed by Shannon \cite{Shannon}. } entropies \cite{IBB2006,Opt,OptMaj} (for more results see the upcoming
review \cite{Cgreview}). At the end of the paper I will go back,
for a moment, to both the continuous and coarse-grained scenario,
while the main discussion will be conducted in the framework of finite-dimensional
quantum systems.

I have devoted the whole paragraph to the description of the landscape
formed by entropic URs, in order to emphasize how fruitful the concept
of Shannon information entropy is. However, to unambiguously achieve
this goal I shall point out that the Shannon and R{\'e}nyi entropies are
useful in various physical contexts, such as: entanglement detection
\cite{entro,Saboia,Reliable}, steering detection \cite{Ster1,Ster1.1,Ster2},
quantum key distribution \cite{QKD}, quantum thermodynamics \cite{ThermoHoro},
resource theories \cite{Plenio} or Bell's inequalities \cite{BrCav,CerfAdami,Bell1,Bell2}
\textemdash{} some without any direct connection to uncertainty
relations (e.g. \cite{ThermoHoro,Plenio,Bell1,Bell2}).

The last topic from the long list presented above, devoted to the
physical underpinnings of the Bell-type experiments, brings the questions
about \emph{reality} of quantum observables. While reality in the
above meaning is usually viewed as a competitor of quantum nonlocality
(due to Bell's theorem quantum mechanics violates Einstein's local realism, so it must either be non-local or not real), thus being a qualitative notion, it is rarely
discussed in a quantitative manner. Dieguez and Angelo \cite{Real2}
phrase that as follows: ``\emph{...too little (if any) has been achieved
with regard to formal connections between elements of reality and
fundamental concepts such as information and quantum correlations.}''
To partially fill this gap a measure of reality of quantum observables
based on information entropies has been proposed in \cite{Real1}.
Deeper links between information and reality (also expressed in a
kind of duality relation) were further investigated in \cite{Real2}.
In particular, it was shown that a completely positive trace preserving
(CPTP) map, called monitoring, increases the reality of every observable
\cite{Real2}.

The goal of the current contribution is to explore connections between
the information-based measures of reality for quantum observables
and entropic measures of uncertainty. The notion of reality, as being
interrelated with the Bell's inequalities and Einstein-Podolsky-Rosen
paradox, must allow for a possibility of entanglement between the
system under consideration and other parties. As a consequence, entropic
URs suitable for the goal assumed are those with, so called, quantum
memory \cite{Berta}. In Sec. \ref{Prel} all necessary tools involving
various entropic measures relevant for bipartite quantum systems are
introduced, while  in Section \ref{Main} the main discussion of the uncertainty-reality complementarity
takes place (appropriate inequalities are derived).  In Section \ref{Diss} we go back to the monitoring quantum channel, as well as to continuous and coarse-grained observables.

\section{Preliminaries}\label{Prel}

Before we move to the main part we need to fix the notation and define
the objects of interest. We consider a mixed state $\varrho_{AB}$
acting on a bipartite Hilbert space $\mathcal{H}_{A}\otimes\mathcal{H}_{B}$.
We make no assumptions about dimensionality of both Hilbert spaces,
denoted by $d_{A/B}=\dim\mathcal{H}_{A/B}$, though in the main part
we will concentrate on finite-dimensional systems. The reduced density
matrix of the second subsystem (after the partial trace) is $\varrho_{B}=\textrm{Tr}_{A}\left(\varrho_{AB}\right)$. 

Let $X$ be a non-degenerate observable acting on $\mathcal{H}_{A}$
and let $\left|x_{i}\right\rangle $ for $i=1,\ldots,d_{A}$ be the
eigenstates of $X$, so that $\sum_{i=1}^{d_{A}}\left|x_{i}\right\rangle \left\langle x_{i}\right|=\1_{A}$.
The post-measurement state is described by the density matrix
\begin{equation}
\varrho_{XB}=\Phi_{X}\left(\varrho_{AB}\right)=\sum_{i=1}^{d_{A}}\left(\left|x_{i}\right\rangle \left\langle x_{i}\right|\otimes\1_{B}\right)\varrho_{AB}\left(\left|x_{i}\right\rangle \left\langle x_{i}\right|\otimes\1_{B}\right).
\end{equation}
The map $\Phi_{X}\left(\cdot\right)$ is completely positive and trace
preserving. In an analogous way we define $\varrho_{YB}$ to be the
post-measurement state relevant for the second observable $Y$, again
acting on $\mathcal{H}_{A}$.

The assumed notation might suggest that the state $\varrho_{XB}$
acts on $\mathcal{H}_{X}\otimes\mathcal{H}_{B}$, and that $\Phi_{X}:\mathcal{H}_{A}\otimes\mathcal{H}_{B}\rightarrow\mathcal{H}_{X}\otimes\mathcal{H}_{B}$,
which in fact is formally consistent (the same applies to $Y$). However,
for simplicity and without loss of generality we set $\mathcal{H}_{X}=\mathcal{H}_{A}=\mathcal{H}_{Y}$.
Still, the first label of the bipartite density matrix encodes the
information whether we deal with the original state ($A$) or post-measurement states ($X$ or $Y$).

We further use several entropic quantities based on the von-Neumann
entropy $S=-\textrm{Tr}\left(\varrho\ln\varrho\right)$ equal to the
Shannon entropy $H$ calculated for the spectrum of the density matrix
in question. In particular, we denote: 
\begin{equation}
H\left(AB\right)=S\left(\varrho_{AB}\right),
\end{equation}
and consequently $H\left(XB\right)=S\left(\varrho_{XB}\right)$. For
the reduced state we employ the notation $H\left(B\right)=S\left(\varrho_{B}\right)$.

Differences of the above quantities form the conditional Shannon (von
Neumann) entropies:\begin{subequations} 
\begin{equation}
H\left(A|B\right)=H\left(AB\right)-H\left(B\right)\equiv S\left(\varrho_{AB}\right)-S\left(\varrho_{B}\right),
\end{equation}
\begin{equation}
H\left(X|B\right)=H\left(XB\right)-H\left(B\right)\equiv S\left(\varrho_{XB}\right)-S\left(\varrho_{B}\right),
\end{equation}
and similarly for $Y$. \end{subequations}  Note that since the state
$\varrho_{AB}$ can, in principle, be entangled, the entropy $H\left(A|B\right)$
can assume negative values. However, the post-measurement state (only
discussed for $X$; the same applies to $Y$) can also be written
as 
\begin{equation}
\varrho_{XB}=\sum_{i=1}^{d_{A}}p_{i}\left|x_{i}\right\rangle \left\langle x_{i}\right|\otimes\sigma_{B}^{(x_{i})},\label{RxbSep}
\end{equation}
with $p_{i}$ and $\sigma_{B}^{(x_{i})}$ being the probabilities
associated with $\left|x_{i}\right\rangle \left\langle x_{i}\right|\otimes\1_{B}$
and the resulting states of the second subsystem respectively. As
$\varrho_{XB}$ in the form (\ref{RxbSep}) is manifestly separable
we know that \cite{separability} $H\left(X|B\right)\geq0$. As a
consequence, $H\left(X|B\right)$ is considered as an entropic notion
of uncertainty in the bipartite setting \cite{Berta}, often referred
to as equipped with quantum memory. Note that if $d_{B}=1$ this quantity
simply reduces to the ordinary Shannon entropy of the probabilities
$p_{i}$, present in all standard entropic URs. 

The memory-assisted entropic UR for the pair of observables $X$ and
$Y$ is of the form \cite{Berta}
\begin{equation}
H\left(X|B\right)+H\left(Y|B\right)\geq q+H\left(A|B\right),\label{memoryUR}
\end{equation}
with $q$ being the MU lower bound for $X$ and $Y$ \cite{MU},
i.e. $q=-2\ln c$ with $c=\max_{i,j}\left|\left\langle x_{i}\left|y_{j}\right\rangle \right.\right|$,
or its improvements \cite{CP}. The major qualitative difference with
respect to the case without quantum memory is due to the presence
of $H\left(A|B\right)$ on the right hand side. For a sufficiently
entangled state the lower bound in (\ref{memoryUR}) is trivial --- equal
to $0$. More importantly, this trivial bound
can be saturated, what means that one can have $H\left(X|B\right)=0=H\left(Y|B\right)$.

We need to leave the notion of uncertainty for a moment, and turn
to the concept of reality. In \cite{Real1,Real2} it was proposed
that reality of an arbitrary observable $X$ results in the equality
$\varrho_{AB}=\varrho_{XB}$, i.e. the post-measurement state is\emph{
the same} as was the state before the measurement. Consequently, one
could quantify the degree of \emph{irreality} of the discussed observable
in terms of distinguishability measures on the space of quantum states.
Again we can observe how influential the contributions of Shannon
are, since probably the most popular choice for the distinguishability
measure (both in classical and quantum scenario) is the (quantum)
relative entropy $D\left(\rho||\sigma\right)$. Due to a special relationship
between original and post-measurement states their relative entropy
assumes a particularly handy form. One thus defines the ``measure''
of irreality of an observable $X$, given the state $\varrho_{AB}$
as \cite{Real1,Real2}
\begin{equation}
\Im\left(X|\varrho_{AB}\right)=D\left(\varrho_{XB}||\varrho_{AB}\right)=S\left(\varrho_{XB}\right)-S\left(\varrho_{AB}\right)\equiv H\left(XB\right)-H\left(AB\right).
\end{equation}
As is the relative entropy, $\Im\left(X|\varrho_{AB}\right)$ is non-negative
and vanishes if and only if the observable in question is real according
to the definition provided above. We shall point out that the definition of $\Im$ does not lead to a distinguished choice for the measure of reality (denoted, e.g. by $\Re$). In particular, there is no obvious argument selecting the form $\Re=\mathrm{Const}-\Im$, with the constant chosen, for instance, as $\log d_A$. However, the "change of reality" happening in any process shall obey the conservation law $\Delta \Re(X)+\Delta \Im(X)=0$ \cite{Real2}. Note that both changes do only depend on the observable $X$, as processes causing the change do in principle affect the quantum state. In Sec.  \ref{Diss} we go back to $\Delta \Re(X)$ while discussing the monitoring map, offering the reader an insight into the physical meaning of the quantities under consideration. 

\section{Uncertainty--Reality Complementarity}\label{Main}

For a single observable $X$ we have two quantities, supposed to grasp
its distinct quantum-mechanical aspects. While $H\left(X|B\right)$
quantifies the uncertainty, $\Im\left(X|\varrho_{AB}\right)$ shall
describe the amount of irreality of $X$. Apart from them, we also
have $H\left(A|B\right)$ which takes care of intrinsic properties
of $\varrho_{AB}$ such as entanglement, being completely independent
from the observable in question.

We immediately observe that these three measures are not independent.
Instead, they are in a basic linear relationship
\begin{equation}
\Im\left(X|\varrho_{AB}\right)=H\left(X|B\right)-H\left(A|B\right).\label{constraint1}
\end{equation}
The above relation does not diminish the meaning of $\Im\left(X|\varrho_{AB}\right)$,
as it merely gives new interpretation to $H\left(A|B\right)$.
If the observable $X$ is less uncertain than irreal, then the state
$\varrho_{AB}$ is necessarily entangled as $H\left(A|B\right)$ is
negative.

If we consider the second observable $Y$ and write down its counterpart
of Eq. \ref{constraint1}, we find that 
\begin{equation}
H\left(X|B\right)-\Im\left(X|\varrho_{AB}\right)=H\left(Y|B\right)-\Im\left(Y|\varrho_{AB}\right),\label{constraint2}
\end{equation}
where both sides of (\ref{constraint2}) are equal to $H\left(A|B\right)$.
As a consequence, the four quantities: $H\left(X|B\right)$, $H\left(Y|B\right)$,
$\Im\left(X|\varrho_{AB}\right)$ and $\Im\left(Y|\varrho_{AB}\right)$
are not mutually independent. Still, every pair of measures, $H\left(X|B\right)$
and $\Im\left(Y|\varrho_{AB}\right)$ for example, or even an arbitrary
triple, are not generically interrelated.

We are in position to write-down inequalities which involve both,
the uncertainty and the irreality. As sometimes happens in the topic
of wave-particle duality\footnote{Note that also here, the information entropies were successfully applied
\cite{WPdual}.}, the results presented below are more a consequence of algebraic
manipulations, rather than deep mathematical theorems used to prove
original entropic URs, such as $L_{p}-L_{q}$ \cite{BBM} or $l_{p}-l_{q}$
\cite{MU} norm inequalities. Still, similarly to the wave-particle duality
problem, the obtained results bring interesting interpretations.

First of all, if we apply Eq. \ref{constraint1} to the uncertainty
relation (\ref{memoryUR}) and utilize (\ref{constraint2}) we get
\begin{equation}
\Im\left(X|\varrho_{AB}\right)+H\left(Y|B\right)=H\left(X|B\right)+\Im\left(Y|\varrho_{AB}\right)\geq q.\label{NewB1}
\end{equation}
The above inequality does only involve the bound $q$ which is state-independent.
This is a substantial qualitative difference with respect to the bound
in (\ref{memoryUR}), which due to its dependence on $\varrho_{AB}$
could be trivial. In other words, in the presence of quantum memory,
the uncertainty relation (\ref{NewB1}) seems to encapsulate the original
meaning of the entropic URs. To substantiate that claim and flash
more light on the \emph{uncertainty\textendash reality complementarity
}expressed by (\ref{NewB1}), we find two additional inequalities
closely related to the former one.

The first inequality is obtained by rewriting (\ref{memoryUR}) in
terms of both measures of irreality. Due to the relation (\ref{constraint1})
and its counterpart for $Y$, not written-down explicitly, we get
\begin{equation}
\Im\left(X|\varrho_{AB}\right)+\Im\left(Y|\varrho_{AB}\right)\geq q-H\left(A|B\right),\label{realityUR}
\end{equation}
which is the uncertainty relation for (ir)reality measures of two
observables. One immediately observes that the lower bounds in (\ref{memoryUR})
and (\ref{realityUR}) differ ``only'' by the sign of the conditional
entropy contribution. This new UR, even though derived directly from
(\ref{memoryUR}), is in a kind of duality relationship with its ancestor.
Entangled states characterized by negative $H\left(A|B\right)$ can
decrease the sum of uncertainties (sometimes trivializing the lower
bound), however, they also result in a joint lack of reality.

To make the observed duality even more visible, we finally take the
sum of (\ref{memoryUR}) and (\ref{realityUR})
\begin{equation}
H\left(X|B\right)+\Im\left(X|\varrho_{AB}\right)+H\left(Y|B\right)+\Im\left(Y|\varrho_{AB}\right)\geq2q.\label{FinalUR}
\end{equation}
We again face the uncertainty relation with state-independent lower
bound. Importantly, the inequality (\ref{FinalUR}) can be saturated
in completely opposite (dual in some sense) cases. To this end we
set $d_{A}=d_{B}\equiv d$, and let $X$ and $Y$ be complementary
(mutually unbiased) observables for which $c=1/\sqrt{d}$, so that
$q=\ln d$. 

For a pure, maximally entangled state we easily find $H\left(AB\right)=0$
as $\varrho_{AB}$ is pure and $H\left(B\right)=q$ because $\varrho_{B}=d^{-1}\1_{B}$
as a consequence of maximal entanglement in $\varrho_{AB}$. Thus
$H\left(A|B\right)=-q$. We can also calculate the post-measurement
state (here for the observable $X$)
\begin{equation}
\varrho_{XB}=\frac{1}{d}\sum_{i=1}^{d}\left|x_{i}\right\rangle \left\langle x_{i}\right|\otimes\left|\overline{x_{i}}\right\rangle \left\langle \overline{x_{i}}\right|,
\end{equation}
with $\left|\overline{x_{i}}\right\rangle $ being the complex conjugate
(coefficientwise) of the ket $\left|x_{i}\right\rangle $ with respect
to the Schmidt basis relevant for $\varrho_{AB}$. Thus $H\left(XB\right)=q$,
and also $H\left(YB\right)=q$ because the state $\varrho_{YB}$ would
clearly have the same eigenvalues ($d$-times degenerate eigenvalue
equal to $1/d$) as $\varrho_{XB}$. Taking all the above partial
results together we get
\begin{equation}\label{pieces}
H\left(X|B\right)=H\left(Y|B\right)=0,\qquad\Im\left(X|\varrho_{AB}\right)=\Im\left(Y|\varrho_{AB}\right)=q.
\end{equation}
We can see that the combined UR (\ref{FinalUR}) is saturated.

As the second example we consider the ``dual'' case in which $\varrho_{AB}=d^{-2}\1_{AB}$
is the maximally mixed state. Clearly $\varrho_{XB}=\varrho_{AB}$
for any $X$, what was already pointed out in \cite{Real1}, so that
both measures $\Im$ are zero. On the contrary, both uncertainties
$H\left(X|B\right)$ and $H\left(Y|B\right)$ are equal to $q$, even
though the reduced state $\varrho_{B}$ is exactly the same as in
the case of the pure, maximally entangled state. Simply, the von Neumann
entropy for all possible post-measurement states is equal to $2q$,
which is the entropy of $\varrho_{AB}$. Again, the lower bound in
(\ref{FinalUR}) becomes saturated, even though uncertainty and irreality swapped their contributions.

\section{Discussion}\label{Diss}

In the last section I sketched a way in which one discovers that the
measure of irreality is complementary to the entropic measure of uncertainty.
Eqs. \ref{NewB1}, \ref{realityUR} and \ref{FinalUR} are to a large
extent equivalent to each other, though each UR conveys a different
physical interpretation. To shortly repeat: in (\ref{NewB1}) we find
the state-independent UR similar in spirit to the seminal results
by Maassen and Uffink; the bound in (\ref{realityUR}) is dual to
the one present in the quantum memory-assisted UR in Eq. \ref{memoryUR};
the UR (\ref{FinalUR}) is the four-term uncertainty relation which
for any pair of observables becomes saturated in both opposite regimes
of pure, maximally entangled states and maximally mixed, fully separable
states.

In \cite{Real2} the CPTP map $\mathcal{M}_{Y}^{\epsilon}$, referred
to as monitoring and defined through its action
\begin{equation}
\mathcal{M}_{Y}^{\epsilon}:\varrho_{AB}\longmapsto\left(1-\epsilon\right)\varrho_{AB}+\epsilon\varrho_{YB},\label{monit}
\end{equation}
has been investigated. A compelling physical intuition behind this map is offered in \cite{Real2}, where interaction with an appropriate ancilla is deeply discussed. For our purpose it is enough to notice that $\left[\mathcal{M}_{Y}^{\epsilon}\right]^n=\mathcal{M}_{Y}^{1-\left(1-\epsilon\right)^n}$, so that several instances of the monitoring map's action do simply increase $\epsilon$, pushing the initial state towards $\varrho_{YB}$.

As already mentioned, the map (\ref{monit}) has extensively been studied in \cite{Real2}, in particular, it was shown that
\begin{equation}
\Im\left(X|\varrho_{AB}\right)\geq\Im\left(X|\mathcal{M}_{Y}^{\epsilon}\left[\varrho_{AB}\right]\right),
\end{equation}
for every pair of observables $X$ and $Y$, so that the monitoring
(with respect to any observable) increases the reality of every other
observable, i.e. $\Delta \Re(X)\geq 0$ for (\ref{monit}).

As $\textrm{Tr}_{A}\left(\mathcal{M}_{Y}^{\epsilon}\left[\varrho_{AB}\right]\right)=\varrho_{B}$,
and $\Phi_{Y}\left(\mathcal{M}_{Y}^{\epsilon}\left[\varrho_{AB}\right]\right)=\varrho_{YB}$,
we can see that the uncertainty of $Y$ for the state $\varrho_{AB}$,
$H\left(Y|B\right)$, remains the same for the monitored state $\mathcal{M}_{Y}^{\epsilon}\left[\varrho_{AB}\right]$.
As a consequence, Eq. \ref{NewB1} written down for $\mathcal{M}_{Y}^{\epsilon}\left[\varrho_{AB}\right]$
reads
\begin{equation}
\Im\left(X|\mathcal{M}_{Y}^{\epsilon}\left[\varrho_{AB}\right]\right)+H\left(Y|B\right)\geq q.\label{BoundMon}
\end{equation}
In other words, irreality of $X$ with respect to the quantum state
monitored by $Y$ is lower-bounded by $q-H\left(Y|B\right)$. For
example, if $X$ and $Y$ are taken to be complementary, i.e. such
that $q=\ln d_{A}$,\textbf{ }the bound is very restrictive as $H\left(Y|B\right)\leq\ln d_{A}$
by construction. In particular, for the monitoring map (\ref{monit}) Eq. \ref{BoundMon} tells us that
\begin{equation}
\Delta \Re(X)=-\Delta \Im(X)=\Im\left(X|\varrho_{AB}\right)-\Im\left(X|\mathcal{M}_{Y}^{\epsilon}\left[\varrho_{AB}\right]\right)\leq \Im\left(X|\varrho_{AB}\right)+H\left(Y|B\right)- q.
\end{equation}
If the state $\varrho_{AB}$ is pure and maximally
entangled, so that $H\left(Y|B\right)=0$, the bound (\ref{BoundMon})
prevents the reality from a severe growth driven by monitoring based
on a "sufficiently complementary" observable. Is the observable $Y$ fully complementary to $X$ (see discussion around Eq. \ref{pieces}) the level of reality remains constant.

At the end, we shall go back to the continuous \cite{BBM} and the
coarse-grained \cite{IBB1984,IBB2006,Opt,OptMaj,Cgreview} scenarios.
Two recent, mathematically oriented contributions \cite{Lieb,Furrer}
extend the URs relevant for the above cases, in order to include quantum-memory
effects. In \cite{Lieb}, a continuous counterpart of Eq. \ref{memoryUR}
\begin{equation}
h\left(x|B\right)+h\left(p|B\right)\geq\ln2\pi+H\left(A|B\right),\label{eq:contin}
\end{equation}
has been derived. By $h\left(x|B\right)$ and $h\left(p|B\right)$
we denote continuous conditional Shannon entropies for position and
momentum probability densities respectively. It is interesting that
in the case without quantum memory, state-independent part of the
lower bound is sharper (equal to $\ln e\pi$ \cite{BBM}), however,
the quantum-memory UR in the form of Eq. \ref{eq:contin} can also
be saturated (i.e. $\ln2\pi$ in Eq. \ref{eq:contin} is optimal) \cite{HallTemp}.

Despite mathematical subtleties related to infinite dimension of the
Hilbert space, one could consider a direct generalization of the major
results of this paper. However, as pointed out in \cite{Furrer} it
is better to define all conditional entropies first in the coarse-grained
setting and then take the fine-graining limit in which the coarse-graining
widths tend to $0$. It might happen that a difference of entropic
measures which diverge in this limit does exist and has a physical
meaning. In this formally safer scenario, however, one finds memory-assisted
entropic URs which differ from (\ref{eq:contin}), mainly because
the involved quantities are also different. Thus, while it would be
interesting to merge the concept of reality (now for continuous and
coarse-grained observables) with entropic measures of uncertainty,
this task requires much more care and attention. For now, it will
be left as a future open problem.

Staying with the questions for further research, one shall point out
that uncertainty in the presence of quantum memory is also well-captured
by the information exclusion principle \cite{Hallexclusion,Grudka,CP},
expressed in terms of  mutual information. All results derived here
could gain additional interpretations, while rewritten that way.

\begin{acknowledgments} 

\L .R. acknowledges financial support by grant number 2015/18/A/ST2/00274
of the National Science Center, Poland. \end{acknowledgments}


\begin{thebibliography}{References}
\bibitem{Shannon} C. Shannon, \emph{The mathematical theory of communication},
Bell System Technical Journal \textbf{27}, 379 (1948).

\bibitem{Filozofia} O. Lombardi, F. Holik, and L. Vanni, \emph{What
is quantum information?}, Stud. Hist. Philos. Mod. Phys. \textbf{56},
17 (2016).

\bibitem{BBM} I. Bia\l ynicki-Birula and J. Mycielski\emph{,} \emph{Uncertainty
relations for information entropy in wave mechanics}, Commun. Math.
Phys. \textbf{44,} 129 (1975).

\bibitem{Deutsch} D. Deutsch\emph{, Uncertainty in quantum measurements},
Phys. Rev. Lett. \textbf{50}, 631 (1983).

\bibitem{MU} H. Maassen and J. B. M. Uffink, \emph{Generalized entropic
uncertainty relations}, Phys. Rev. Lett. \textbf{60}, 1103 (1988).

\bibitem{Ozawa}  F. Buscemi, M. J. W. Hall, M. Ozawa, and M. M. Wilde,
\emph{Noise and Disturbance in Quantum Measurements: An Information-Theoretic
Approach}, Phys. Rev. Lett. \textbf{112}, 050401 (2014).

\bibitem{SRuiz} J. I. de Vicente and J. S{\'a}nchez-Ruiz, \emph{Improved
bounds on entropic uncertainty relations}, Phys. Rev. A \textbf{77},
042110 (2008).

\bibitem{my1} Z. Pucha\l a, \L . Rudnicki, and K. \.{Z}yczkowski,
\emph{Majorization entropic uncertainty relations}, J. Phys. A: Math.
Theor. \textbf{46}, 272002 (2013).

\bibitem{Oni} S. Friedland, V. Gheorghiu, and G. Gour, \emph{Universal
Uncertainty Relations}, Phys. Rev. Lett. \textbf{111}, 230401 (2013).

\bibitem{CP} P. Coles and M. Piani, \emph{Improved entropic uncertainty
relations and information exclusion relations}, Phys. Rev. A \textbf{89},
022112 (2014).

\bibitem{my2} \L . Rudnicki, Z. Pucha\l a, and K. \.{Z}yczkowski,
\emph{Strong majorization entropic uncertainty relations}, Phys. Rev.
A \textbf{89}, 052115 (2014).

\bibitem{KamilK} K. Korzekwa, M. Lostaglio, D. Jennings, and T. Rudolph,
\emph{Quantum and classical entropic uncertainty relations}, Phys.
Rev. A \textbf{89}, 042122 (2014). 

\bibitem{Zozor} S. Zozor, G. M. Bosyk, and M. Portesi, \emph{General
entropy-like uncertainty relations in finite dimensions}, J. Phys.
A:Math. Theor. \textbf{47}, 495302 (2014). 

\bibitem{Kaniewski} J. Kaniewski, M. Tomamichel, and S.Wehner, \emph{Entropic
uncertainty from effective anticommutators}, Phys. Rev. A \textbf{90},
012332 (2014). 

\bibitem{my3} Z. Pucha\l a, \L . Rudnicki, A. Krawiec, and K. \.{Z}yczkowski,\emph{
Majorization uncertainty relations for mixed quantum states}, J. Phys.
A: Math. Theor. \textbf{51}, 175306 (2018).

\bibitem{Wehner2}  P. J. Coles, M. Berta, M. Tomamichel, and S. Wehner,
\emph{Entropic Uncertainty Relations and their Applications}, Rev.
Mod. Phys. \textbf{89}, 015002 (2017).

\bibitem{Wehner} S. Wehner and A. Winter, \emph{Entropic uncertainty
relations\textemdash a survey}, New J. Phys. \textbf{12}, 025009 (2010).

\bibitem{Sen} I. Bialynicki-Birula, and \L . Rudnicki, \emph{Entropic
uncertainty relations in quantum physics} in: \textquotedbl{}\emph{Statistical
Complexity: Applications in Electronic Structure}\textquotedbl{},
K. D. Sen (ed.), Springer, p. 1-34 (2011).

\bibitem{Partovi}  M. H. Partovi, \emph{Entropic Formulation of Uncertainty
for Quantum Measurements}, Phys. Rev. Lett. \textbf{50}, 1883 (1983).

\bibitem{IBB1984} I. Bialynicki-Birula, \emph{Entropic uncertainty
relations}, Phys. Lett. \textbf{103}A, 253 (1984). 

\bibitem{IBB2006} I. Bialynicki-Birula, \emph{Formulation of the
uncertainty relations in terms of the R{\'e}nyi entropies}, Phys. Rev.
A \textbf{74}, 052101 (2006).

\bibitem{Opt} \L . Rudnicki, S. P. Walborn, and F. Toscano, \emph{Optimal
uncertainty relations for extremely coarse-grained measurements},
Phys. Rev. A \textbf{85}, 042115 (2012).

\bibitem{OptMaj} \L . Rudnicki, \emph{Majorization approach to entropic
uncertainty relations for coarse-grained observables}, Phys. Rev.
A \textbf{91}, 032123 (2015). 

\bibitem{Cgreview} F. Toscano, D. S. Tasca, \L . Rudnicki, and S.
P. Walborn, \emph{Uncertainty relations for Coarse-Grained Measurements:
an Overview}, in preparation (2018).

\bibitem{entro} S. P. Walborn, B. G. Taketani, A. Salles, F. Toscano,
and R. L. de Matos Filho, \emph{Entropic Entanglement Criteria for
Continuous Variables}, Phys. Rev. Lett. \textbf{103}, 160505 (2009).

\bibitem{Saboia} A. Saboia, F. Toscano, and S. P. Walborn, \emph{Family
of continuous-variable entanglement criteria using general entropy
functions}, Phys. Rev. A \textbf{83}, 032307 (2011).

\bibitem{Reliable} D. S. Tasca, \L . Rudnicki, R. M. Gomes, F. Toscano,
and S. P. Walborn, \emph{Reliable Entanglement Detection Under Coarse\textendash Grained
Measurements}, Phys. Rev. Lett. \textbf{110}, 210502 (2013).

\bibitem{Ster1} J. Schneeloch, P. B. Dixon, G. A. Howland, C. J.
Broadbent, and J. C. Howell, \emph{Violation of continuous-variable
Einstein-Podolsky-Rosen steering with discrete measurements}, Phys.
Rev. Lett. \textbf{110}, 130407 (2013). 

\bibitem{Ster1.1} J. Schneeloch, C. J. Broadbent, S. P. Walborn,
E. G. Cavalcanti, and J. C. Howell, \emph{EPR Steering Inequalities
from Entropic Uncertainty Relations}, Phys. Rev. A \textbf{87}, 062103
(2013).

\bibitem{Ster2} J. Schneeloch, C. J. Broadbent, and J. C. Howell,
\emph{Improving Einstein\textendash Podolsky\textendash Rosen steering
inequalities with state information}, Phys. Lett. A \textbf{378},
766 (2014).

\bibitem{QKD} M. Koashi, \emph{Unconditional security of quantum
key distribution and the uncertainty principle}, J. Phys.: Conf. Ser.
\textbf{36}, 98 (2006).

\bibitem{ThermoHoro} F. Brand\~{a}o, M. Horodecki, N. Ng, J. Oppenheim,
and S. Wehner, \emph{The second laws of quantum thermodynamics}, PNAS
\textbf{112}, 3275 (2015).

\bibitem{Plenio} T. Baumgratz, M. Cramer, and M. B. Plenio, \emph{Quantifying
Coherence}, Phys. Rev. Lett. \textbf{113}, 140401 (2014).

\bibitem{BrCav} S. L. Braunstein and C. M. Caves, \emph{Information-Theoretic Bell Inequalities}, Phys.Rev. Lett. \textbf{61}, 662  (1988). 

\bibitem{CerfAdami} N. J. Cerf and C. Adami, \emph{Entropic Bell Inequalities}, Phys.Rev. A \textbf{55}, 3371  (1997).

\bibitem{Bell1} R. Chaves, L. Luft, and D. Gross, \emph{Causal structures
from entropic information: Geometry and novel scenarios}, New J. Phys.
\textbf{16}, 043001 (2014). 

\bibitem{Bell2} R. Chaves and C. Budroni, \emph{Entropic nonsignalling
correlations}, Phys. Rev. Lett. \textbf{116}, 240501 (2016).

\bibitem{Real2} P. R. Dieguez and R. M. Angelo, \emph{Information-reality
complementarity: The role of measurements and quantum reference frames},
Phys. Rev. A \textbf{97}, 022107 (2018).

\bibitem{Real1} A. L. O. Bilobran and R. M. Angelo, \emph{A measure
of physical reality}, Europhys. Lett. \textbf{112}, 40005 (2015).

\bibitem{Berta} M. Berta, M. Christandl, R. Colbeck, J. M. Renes,
and R. Renner, \emph{The uncertainty principle in the presence of
quantum memory}, Nature Phys. \textbf{6}, 659 (2010).

\bibitem{separability} R. Horodecki and M. Horodecki, \emph{Information-theoretic
aspects of inseparability of mixed states}, Phys. Rev. A \textbf{54},
1838 (1996).

\bibitem{WPdual} P. J. Coles, J. Kaniewski, and S. Wehner, \emph{Equivalence
of wave-particle duality to entropic uncertainty}, Nat. Comm. \textbf{5},
5814 (2014).

\bibitem{Lieb} R. L. Frank and E. H. Lieb, \emph{Extended Quantum
Conditional Entropy and Quantum Uncertainty Inequalities}, Commun.
Math. Phys. \textbf{323,} 487 (2013).

\bibitem{Furrer} F. Furrer, M. Berta, M. Tomamichel, V. B. Scholz,
and M. Christandl, \emph{Position-momentum uncertainty relations in
the presence of quantum memory}, J. Math. Phys. \textbf{55}, 122205
(2014).

\bibitem{HallTemp} M. J. W. Hall, \emph{Universal geometric approach to uncertainty, entropy, and information}, Phys. Rev. A \textbf{59},
2602 (1999).

\bibitem{Hallexclusion} M. J. W. Hall, \emph{Information Exclusion
Principle for Complementary Observables}, Phys. Rev. Lett. \textbf{74},
3307 (1995).

\bibitem{Grudka} A. Grudka, M. Horodecki, P. Horodecki, R. Horodecki,
W. K\l obus, and \L . Pankowski, \emph{Conjectured strong complementary-correlations
tradeoff}, Phys. Rev. A \textbf{88}, 032106 (2013).
\end{thebibliography}
\end{document}